\documentclass[11pt,letterpaper,twoside]{article}
\usepackage[centertags]{amsmath}
\usepackage{graphicx,indentfirst,amsmath,amsfonts,amssymb,amsthm,newlfont}
\font\Goth=yinitas scaled \magstep0

\newcommand{\Gth}[1]{\lower2mm\hbox{\Goth #1}}

\def\al{\alpha}

\def\eps{\varepsilon}

\def\de{\delta}

\def\be{\beta}

\def\l1{{\lambda}_1}

\newcommand{\f}{\frac}

\def\ln{\mbox{\rm ln}}

\def\x1{{\xi }_{xx}}
\def\x2{{\xi }_{yy}}
\def\x3{{\xi }_{xy}}

\def\e1{{\eta }_{xx}}
\def\e2{{\eta }_{yy}}
\def\e3{{\eta }_{xy}}

\newcommand{\ds}{\displaystyle }

\newcommand{\beqn}{\begin{eqnarray*}}
\newcommand{\eeqn}{\end{eqnarray*}}
\newcommand{\beqnn}{\begin{eqnarray}}
\newcommand{\eeqnn}{\end{eqnarray}}
\newcommand{\p}{\partial}
\newcommand{\bb}{\begin{equation}}
\newcommand{\ee}{\end{equation}}
\newcommand{\ba}{\begin{array}}
\newcommand{\ea}{\end{array}}
\newcommand{\R}{\mathbb{R}}

\begin{document}
\pagenumbering{arabic}
\title{\huge \bf Strict self-adjointness and certain shallow water models}
\author{\rm \large Priscila Leal da Silva and Igor Leite Freire \\
\\
\it Centro de Matem\'atica, Computa\c c\~ao e Cogni\c c\~ao\\ \it Universidade Federal do ABC - UFABC\\ \it 
Rua Santa Ad\'elia, $166$, Bairro Bangu,
$09.210-170$\\\it Santo Andr\'e, SP - Brasil\\
\rm E-mail: igor.freire@ufabc.edu.br/igor.leite.freire@gmail.com\\
\rm E-mail: priscila.silva@ufabc.edu.br/pri.leal.silva@gmail.com}
\date{\ }
\maketitle
\vspace{1cm}
\begin{abstract}
We consider a class of third order equations from the point of view of strict self-adjointness. Necessary and sufficient conditions for the investigated class to be strictly self-adjoint are obtained. Then, from a strictly self-adjoint subclass we consider those who admit a suitable scaling transformation. Consequently, a family of equations including the Benjamin-Bona-Mahony, Camassa-Holm and Novikov equations is deduced. By a suitable choice of the parameters, we deduce an one-parameter family of equations unifying the last two mentioned equations. Then, using some recent techniques for constructing conserved vectors, we show that from the scale invariance it is obtained, as a conserved density, the same quantity employed to construct one of the well known Hamiltonians for the cited integrable equations.
\end{abstract}
\vskip 1cm
\begin{center}
{2010 AMS Mathematics Classification numbers:\vspace{0.2cm}\\
76M60, 58J70, 35A30, 70G65\vspace{0.2cm} }
\end{center}
\pagenumbering{arabic}
\newpage

\section{Introduction}

Since the celebrated Korteweg and de Vries paper \cite{kdv}, in which a third order evolution equation was derived and named after them, a huge number of papers in the literature has been done for modeling, or related with, shallow water equations. During the last century, a sequence of papers, starting with \cite{miu1}, showed and enlightened many properties of such equation. Additionally, the KdV equation
\bb\label{1.1}
u_{t}=u_{xxx}+uu_{x} 
\ee 
proved to be a prototype equation for many phenomena, see, for instance, \cite{ablo}.

Although its good and versatile properties, the equation was not above criticisms. In the seminal paper \cite{bbm}, the authors derived a new equation for moderately long wave equations of small amplitudes whose formal justification is as that for the KdV and from that paper arose the well know Benjamin-Bona-Mahoney (BBM) equation 
\bb\label{1.2}
u_{t}=u_{txx}+uu_{x}.
\ee
However, the differences between both equations are greater than the fact that (\ref{1.1}) is an evolution equation whereas (\ref{1.2}) is not. In \cite{bbm} the authors found three conserved quantities on the solutions of (\ref{1.2}). Later in \cite{ol2}, those obtained conservation laws were proved to be the only three admitted by (\ref{1.2}). This fact shows a dramatic difference between (\ref{1.2}) and (\ref{1.1}) since the first one admits an infinite number of conserved quantities \cite{miu2}. 

More recently, Camassa and Holm \cite{camassa} using Hamiltonian methods derived the famous Camassa-Holm (CH) equation
\bb\label{1.3}
u_{t}-u_{txx}+3uu_{x}=2u_{x}u_{xx}+uu_{xxx}.
\ee
The last equation possesses remarkable properties such as solutions with peaks in which the first order derivatives are discontinuous, called {\it peakon} solutions, and it has a bi-hamiltoninan structure, see \cite{camassa}, which implies in the existence of an infinite number of conserved quantities, like the KdV equation \cite{gard,miu2,miu3}.

Since then, a considerable number of papers have been dedicated to derive third order non-evolutionary dispersive equations having similar properties as those known to KdV and CH equation. To cite a few number of examples, it was derived in \cite{dehoho} an integrable equation having peakon solutions with first order nonlinearities, while in \cite{dugoho} another integrable equation, combining linear dispersion such as the KdV equation and a nonlinear dispersion like the CH equation, was discovered. More recently, Novikov \cite{nov} has discovered the equation
\bb\label{1.4}
u_{t}-u_{txx}+4u^2 u_{x}=3uu_{x}u_{xx}+u^{2}u_{xxx},
\ee
which not only admits peakon solutions and cubic nonlinearities, but it is also integrable \cite{how}.

In \cite{miu2} it was shown that the KdV equation possesses infinitely many conservation laws. This was the start point of a considerable number of papers dealing with the properties of a certain equation and the existence of an infinite number of conserved quantities. Then we arrive at the point of integrability and the existence of infinitely many conservation laws of an equation.

Noether theorem showed a deeper and closer relation between symmetries and conservation laws for the Euler-Lagrange equations. From Noether theorem, for each Noether symmetry of the Euler-Lagrange equation, one can establish a conservation law. Although the KdV equation is not an Euler-Lagrange equation, it can be transformed in one.

Therefore a question naturally arise: would it be possible to derive infinitely many conserved quantities of the KdV equation from its symmetries? This question does not make sense if it is restricted to Lie point symmetries since the KdV equation (or its ``variational form'') admits a finite dimensional symmetry Lie algebra and, therefore, one would expect no more than a finite number of conservation laws coming from Noether's theorem. The infinite number of conservation laws of the KdV equation could be interpreted as the existence of {\it higher order symmetries}, see \cite{ol1}.

And, up to our knownlegement, the first paper relating symmetries (not necessarily Lie point symmetries) of the KdV equation and an infinite number of local conservation laws for it was \cite{ib2}, in which Ibragimov showed how to construct local conservation laws using symmetries other than the Lie point symmetries.

In order to construct the conserved vectors, Ibragimov first established a non-local conserved vector. Then he showed that the KdV equation is strictly self-adjoint \cite{ib1,ib2,ib8} and, consequently, the non-local conserved quantities can be transformed in locals one. These concepts will be better discussed in section \ref{conceitos}.

A considerable number of integrable equations has this common property: strict self-adjointness. In fact, Ibragimov \cite{ib2} showed that KdV is strictly self-adjoint. In \cite{ib7} it was shown that the CH equation has also the same property, as well as in \cite{bfi} it was proved that the Novikov equation is strictly self-adjoint. In particular, with respect to (\ref{1.3}) and (\ref{1.4}), the obtained results in \cite{clark}, \cite{ib7} and \cite{bfi} shows some common facts:
\begin{enumerate}
\item both equations are strictly self-adjoint;
\item both equations admit the scaling symmetry $(x,t,u)\mapsto ( x,\lambda^{-b} t,\lambda u)$, for a certain value of $b$, whose corresponding generator is
\bb\label{1.5}
X_{b}=u\f{\p}{\p u}-bt\f{\p}{\p t};
\ee
\item from the Lie point symmetry generator (\ref{1.5}) and the results proposed in \cite{ib2}, it was obtained the conserved density $u^2+\eps u_{x}^2$ for both equations, see \cite{ib7,bfi}. In fact, let $m=u-\eps u_{xx}$ and assume that $u(x,t)\rightarrow0$ when $x\rightarrow\pm\infty$. Taking the expression $(uu_{x})_{x}=u_{x}^2+uu_{xx}$ into account, it is easily concluded that
$$H_{1}=\int_{-\infty}^{+\infty}m udx=\int_{-\infty}^{+\infty}(u^2+\eps u_{x}^2)dx.$$
Such conserved quantity provides a Hamiltonian to CH \cite{camassa}, Dullin-Gotwald-Holm (DGH) \cite{dugoho} and Novikov \cite{how} equations.
\end{enumerate}

Since Ibragimov's concepts on self-adjointness \cite{ib2,ib3,ib6,ib8} have been introduced, a considerable number of papers have been dealing with the problem of finding classes of differential equations with some self-adjoint property, see, for instance, \cite{ijnmp,iamc,ijjpa,icam,icnsns,gandjpa,totr}. 

Therefore, motivated by those recent results and provoked by the classification carried out in \cite{nov}, in which certain generalizations of the CH equation possessing infinite hierarchies of higher symmetries were considered, in this paper we determine which conditions are necessary and sufficient for the equation
\bb\label{1.6}
u_{t}+\eps u_{txx}+f(u)u_{x}+g(u)u_{x}u_{xx}+h(u)u_{xxx}=0
\ee
to be strictly self-adjoint. 

Once having carried out the strict self-adjointness classification of (\ref{1.6}), we restrict ourselves to find the subfamily admitting the scale invariance $(x,t,u)\mapsto ( x,\lambda^{-b} t,\lambda u)$. It is then obtained the following fourth-parameter family of strictly self-adjoint equations 
\bb\label{1.7}
u_{t}+\eps u_{txx}+\gamma u^{b}u_{x}=(b+1)\be u^{b-1}u_{x}u_{xx}+\be u^{b}u_{xxx},
\ee
which includes equations (\ref{1.2}), (\ref{1.3}) and (\ref{1.4}). Moreover, taking $b=1,\,\eps=-\be=\al^2$ and $\gamma=3$, we arrive, up to a translation $u\mapsto u+u_{0}/\al^2$, at the DGH equation 
\bb\label{1.8}
u_{t}-\al^2u_{txx}+3uu_{x}=\al^{2}(uu_{xxx}+2u_{x}u_{xx})+u_{0}u_{xxx},
\ee
which is also integrable, see \cite{dugoho}. The term $u_{0}$ corresponds to the coefficient of the linear dispersion of the equation and when $u_{0}\rightarrow0$ and $\al=1$, such equation turns back to the CH equation. However, if $u_{0}\neq0$, (\ref{1.8}) does not admit the generator (\ref{1.5}). We observe that at the limit of the dispersionless $u_{0},\,\al\rightarrow0$, equation (\ref{1.8}) is reduced to the Riemann equation $u_{t}+3uu_{x}=0$. More generally, when the dispersion effects are neglected in (\ref{1.7}), that is, $\eps,\,\be\rightarrow0$, one obtains a family of Riemann equations given by $u_{t}+\gamma u^{b}u_{x}=0$.

Finally, if we choose $\eps=-1$ and $\gamma=\be(b+2)$, equation (\ref{1.7}) can be rewritten as
\bb\label{1.9}
u_{t}-u_{txx}+\be(b+2)u^{b}u_{x}=(b+1)\be u^{b-1}u_{x}u_{xx}+\be u^{b}u_{xxx}.
\ee
Therefore, defining $m=u-u_{xx}$, (\ref{1.9}) is equivalent to 
\bb\label{1.10}
m_{t}-\be u^{b}m_{x}-\be(b+1)u^{b-1}u_{x}m=0.
\ee
Equation (\ref{1.9}), or its equivalent form (\ref{1.10}), still contains the CH (\ref{1.3}) and Novikov (\ref{1.4}) equations, but not BBM (\ref{1.2}).

Local conserved currents for the family of strictly self-adjoint scale-invariant equations found are presented in section \ref{leis}. Finally, in section \ref{dis} we discuss the obtained results. In section \ref{conceitos} we present some basic facts about Lie symmetries, conservation laws and strict self-adjointness. Then, in section  \ref{classification} it is obtained the family (\ref{1.7}) from (\ref{1.6}).

\section{Lie point symmetries}\label{symmetries}

Here we present some basic facts regarding Lie symmetries of differential equations. In what follows, we shall only consider single partial differential equations, however, the discussed theory can be used for ordinary differential equations and also for systems of differential equations. Further and better discussions on this subject can be found in \cite{ba,bk,i,i1,ol5}.

Since in our main problem we shall consider time dependent equations, in what follows such variable will be represented with either symbols $x^0$ or $t$, depending on the situation. Let $x=(x^{1},\cdots,x^{n})\in X\subseteq\R^n$, $u=u(t,x)\in U\subseteq\R$ and $u_{(j)}$ be, respectively, $n$ independent and a dependent variable, while the set of all $jth$ derivatives of $u$. Hereafter, the summation over repeated indices is presupposed. All functions here are assumed to be smooth. In particular, $u_{i_{1}\cdots i_{j}}=D_{i_{1}}\cdots D_{i_{j}}(u),$
where 
$$D_{i}=\f{\p}{\p x^{i}}+u_{i}\f{\p}{\p u}+u_{ij}\f{\p}{\p u_{j}}+\cdots,\,\,\,\,i=0,\cdots,n$$
are the total derivative operators.

Let ${\cal A}$ be the set of all locally analytic functions of a finite number of the variables $x,\,u$ and $u_{(j)}$. Let $F\in{\cal A}$ and consider an equation
\bb\label{2.1}
F(x,u,u_{(1)},\cdots,u_{(k)})=0.
\ee

Let $M=X\times U\approx\R^{n+1}$ be the space of the independent and dependent variables and consider a local group of transformations $G$ acting on an open subset of $M$. Then the action on such space induces a local action on the jet space $(x,u,u_{(1)},u_{(2)},\cdots,u_{(k)})$, called $k-$jet space. For further details, see \cite{ol5}, Chapter 2.

Let
\bb\label{s1}
X=\xi^{i}(x,u)\f{\p}{\p x^i}+\eta(x,u)\f{\p}{\p u}
\ee
be a generator of the local one-parameter group $ \eps\mapsto\exp{\eps X}(x,u)$ acting on $M$. One says that the $X$ is a Lie point symmetry generator of the equation (\ref{2.1}) if
\bb\label{s2}
X^{(k)}F=\lambda F,
\ee
for a certain function $\lambda$ depending on $x,u,u_{(1)},\cdots$.
Equation (\ref{s2}) is called invariance condition and
\bb\label{s3}
X^{(k)}=\xi^{i}(x,u)\f{\p}{\p x^i}+\eta(x,u)\f{\p}{\p u}+\zeta_{i}\f{\p}{\p u_{i}}+\zeta_{ij}\f{\p}{\p u_{ij}}+\cdots+\zeta_{i_{1}\cdots i_{k}}\f{\p}{\p u_{i_{1}\cdots i_{k}}},
\ee
where
$$\zeta_{i}=D_{i}(\eta)-D_{i}\xi^j u_{j},\cdots,\zeta_{i_{1}\cdots i_{k}}=D_{i_{1}}\cdots D_{i_{1}}(\eta)-D_{i_{k}}\xi^j u_{i_{1}\cdots i_{k-1}j},$$
is the $k-th$ prolongation of the vector field $X$. In this case, we say that 
$$(x,u)\mapsto \exp{\eps X}(x,u):=\left(x+\eps X(x)+\f{\eps^2}{2}X(X(x))+\cdots,u+\eps X(u)+\f{\eps^2}{2}X(X(u))+\cdots\right)$$
is a Lie point symmetry of (\ref{2.1}).

Consider equation (\ref{1.6}) and the transformation 
\bb\label{s4}
(x,t,u)\mapsto ( x,\lambda^{-b} t,\lambda u),\,\lambda>0.
\ee
Let us determine for which functions $f=f(u),\,g=g(u)$ and $h=h(u)$ such transformation is a Lie point symmetry of (\ref{1.6}).

Firstly we observe that (\ref{1.5}) is the generator of the transformation (\ref{s4}). Then, its third extension is
\begin{eqnarray*}
X^{(3)}=u\f{\p}{\p u}-bt\f{\p}{\p t}+(b+1)u_{t}\f{\p}{\p u_{t}}+u_{x}\f{\p}{\p u_{x}}+u_{xx}\f{\p}{\p u_{xx}}+u_{xxx}\f{\p}{\p u_{xxx}}+(b+1)u_{txx}\f{\p}{\p u_{txx}}.
\end{eqnarray*}

A simple calculation shows that
\begin{eqnarray*}
X^{(3)}(u_{t}+\eps u_{txx}&+&f(u)u_{x}+g(u)u_{x}u_{xx}+h(u)u_{xxx})= \\&&(b+1)(u_{t}+\eps u_{txx})+(uf)'u_{x} +[(ug)'+g]u_{x}u_{xx}+(uh)'u_{xxx},
\end{eqnarray*}
where the prime $'$ means derivative with respect to $u$. Therefore, condition (\ref{s2}) gives
\begin{eqnarray*}
(b+1)(u_{t}+\eps u_{txx})&+&(uf)'u_{x}+[(ug)'+g]u_{x}u_{xx}+(uh)'u_{xxx}= \\ && \lambda (u_{t}+\eps u_{txx}+f(u)u_{x}+g(u)u_{x}u_{xx}+h(u)u_{xxx}).
\end{eqnarray*}

From the coefficients of $u_{t},\,u_{x},\,u_{x}u_{xx}$ and $u_{xxx}$ we respectively obtain:
$$
\lambda=b+1,\,\,(uf)'=\lambda f,\,(ug)'+g=\lambda g,\,(uh)'=\lambda h,
$$
which reads $f(u)=\gamma u^{b},\,g(u)=\sigma u^{b-1}$ and $h(u)=\delta u^{b}$, where $\gamma,\,\sigma$ and $\delta$ are arbitrary constants. Then we conclude that the transformation (\ref{s4}) is a Lie point symmetry of the equation (\ref{1.6}) if and only if the equation takes the form
\bb\label{1.6'}
u_{t}+\eps u_{txx}+\gamma u^{b}u_{x}+\sigma u^{b-1}u_{x}u_{xx}+\delta u^{b}u_{xxx}=0.
\ee

\section{Conservation laws obtained from point symmetries}\label{conceitos}

Here we present some elements regarding conservation laws. However, the interested reader is refereed to \cite{ab,ib2,ib6,ib8,ol2,ol3,ol4,vin} for further details. We also guide the curious reader to \cite{i1,natalia,ol5,pop1,pop2,pop3} for additional readings.

\subsection{Conservation laws}

Mathematically speaking, one can define a conservation law for (\ref{2.1}) starting from the expression
\bb\label{2.2}
Div(C):=D_{t}C^0+D_{i}C^i=\lambda F,
\ee
for a certain vector field $C:=(C^0,C_{x})$, where $C_{x}:=(C^1,\cdots, C^n)$, and functions $\lambda=\lambda(t,x,u,\cdots)$. Equation (\ref{2.2}) is called characteristic form of the conservation law $D_{t}C^0+D_{i}C^i=0$, while $\lambda$ is the characteristic of it.

A vector field $C=(C^0,C_{x})$ provides a {\it trivial conservation laws} if $Div(C)\equiv0$. Such a vector $C$ is, therefore, called {\it trivial conserved vector}. Otherwise $C$ is called {\it nontrivial} conserved vector\footnote{A more general discussion can be found in \cite{natalia}. Here we employ such definition of trivial conserved vector because it is enough for our purposes, mainly because we will use it in Section 5.}. 

If $C$ is a trivial conserved vector, for any differential equation (\ref{2.1}), equation (\ref{2.2}) holds with the characteristic $\lambda=0$. Two conserved vectors are said to be {\it equivalent} if they differ by a trivial conserved vector. Clearly two equivalent conserved vectors possess the same characteristic $\lambda$. A conservation law of (\ref{2.2}) can now be rigorously defined as follows.

By {\it conservation law} of (\ref{2.1}) we mean the equivalence class of conserved vectors of (\ref{2.1}). Then, the set of all conservation laws is a vector space whose the identity is the equivalence class of the trivial conserved quantities.

On $(\ref{2.1})$, the relation (\ref{2.2}) becomes $D_{t}C^0+D_{i}C^i\equiv0$. From the physical point of view, the vector field $C=(C^0,C^1,\cdots, C^n)$ is usually a {\it density} and it is called {\it conserved vector or conserved current} of the modeled phenomena by (\ref{2.1}). The component $C^0$ is the conserved density while the remaining components are the conserved flux. Being a density, restricting $x$ to a fixed domain $\Omega\subseteq\R^n$, with a smooth, constant, boundary $\p\Omega$, and defining 
$$Q_{\Omega}=\int_{\Omega}C^{0}dx,$$
application of the divergence theorem gives
$$
\f{d Q_{\Omega}}{dt}=\int_{\Omega}D_{t}C^0dx=-\int_{\Omega}D_{i}C^idx=-\int_{\p\Omega}C_{x}\cdot dS.
$$ 
 Therefore, restricted to $\Omega$, the quantity $Q_{\Omega}$ depends only on the behavior of the solutions on the boundary $\p\Omega$ and it is equal to the total flux over it. For non-dissipative physical model, this fact provides the general form of a conservation law.
 
\subsection{How to construct local conserved currents for an equation}

Although its importance, the construction of conserved quantities is a sensitive point. There are a lot of techniques for dealing with this matter \cite{ab,i1,ib2,ol5,vin}. In this paper we will use the ideas introduced by Ibragimov in \cite{ib2} for constructing local conserved currents for equations of the type (\ref{1.6}).

Firstly, let (\ref{2.1}) be a given differential equation, ${\cal L}:=vF$ be another differential function, called {\it formal Lagrangian}, and 
$$
\f{\de}{\de w}=\f{\p}{\p w}+\sum_{j=0}^{\infty}(-1)^{j}D_{i_{1}}\cdots D_{i_{j}}\f{\p}{\p w_{i_{1}\cdots i_{j}}}
$$
be the Euler-Lagrange operator. Taking the formal Lagrangian into account, from the Euler-Lagrange equations
$\de{\cal L}/\de u=0,\,\,\,\de{\cal L}/\de v=0$,
it is obtained the system $F=0$ and $F^{\ast}=0$, where
\bb\label{2.2.2}
F^{\ast}:=\f{\de{\cal L}}{\de u}=0
\ee
is the adjoint equation to (\ref{2.1}).

In \cite{ib2} Ibragimov showed that if 
\bb\label{2.3}
X=\xi^i(x,u)\f{\p}{\p x^i}+\eta^{\al}(x,u)\f{\p}{\p u}
\ee
is a Lie point symmetry generator of (\ref{2.1}), then
\bb\label{2.8}
\ba{lcl}
C^{i}&=&\ds{\xi^{i}{\cal L}+W\,\left[\f{\p{\cal L}}{\p u_{i}}-D_{j}\left(\f{\p{\cal L}}{\p u_{ij}}\right)-D_{j}D_{k}\f{\p{\cal L}}{\p u_{ijk}}-\cdots\right]}\\
\\
&&\ds{+D_{j}(W)\,\left[\f{\p{\cal L}}{\p u_{ij}}-D_{k}\left(\f{\p{\cal L}}{\p u_{ijk}}\right)+\cdots\right]+D_{j}D_{k}(W)\,\left[\f{\p{\cal L}}{\p u_{ijk}}-\cdots\right]+\cdots},
\ea
\ee
where $W=\eta-\xi^j u_{j}$, provides a conserved vector for the system formed by (\ref{2.1}) and (\ref{2.2.2}). We guide the interested reader to \cite{ib8} for a better and deeper discussion about this subject.

 In particular, for equations of the type (\ref{1.6}) admitting a Lie point symmetry generator
$$X=\tau\f{\p}{\p t}+\xi\f{\p}{\p x}+\eta\f{\p}{\p u},$$
 the components (\ref{2.8}) become
\bb\label{3.4}
\ba{lcl}
C^{0}&=&\ds{\tau{\cal L}+W\left[\f{\p{\cal L}}{\p u_{t}}+D_{x}^2\left(\f{\p{\cal L}}{\p u_{txx}}\right)\right]-D_{x}(W)D_{x}\left(\f{\p{\cal L}}{\p u_{txx}}\right)+D_{x}^{2}(W)\f{\p{\cal L}}{\p u_{txx}}},\\
\\
C^{1}&=&\ds{\xi{\cal L}+W\left[\f{\p{\cal L}}{\p u_{x}}- D_{x}\left(\f{\p {\cal L}}{\p u_{xx}}\right)+D_{x}^2\left(\f{\p{\cal L}}{\p u_{xxx}}\right)+D_{x}D_{t}\left(\f{\p{\cal L}}{\p u_{xxt}}\right)+D_{t}D_{x}\left(\f{\p{\cal L}}{\p u_{xtx}}\right)\right]}\\
\\
&&\ds{-D_{x}(W)D_{x}\left[\f{\p{\cal L}}{\p u_{xx}}-D_{x}\left(\f{\p{\cal L}}{\p u_{xxx}}\right)-D_{t}\left(\f{\p{\cal L}}{\p u_{xxt}}\right)\right]-D_{t}(W)D_{x}\left(\f{\p{\cal L}}{\p u_{xtx}}\right)+D_{x}^{2}(W)\f{\p{\cal L}}{\p u_{xxx}}}\\
\\
&&\ds{+D_{t}D_{x}(W)\f{\p{\cal L}}{\p u_{xtx}}+D_{x}D_{t}(W)\f{\p{\cal L}}{\p u_{xxt}}},
\ea
\ee
where the formal Lagrangian is given by
\bb\label{3.5}
{\cal L}=v\left[u_{t}+\eps \f{u_{txx}+u_{xtx}+u_{xxt}}{3} +f(u)u_{x}+g(u)u_{x}u_{xx}+h(u)u_{xxx}\right].
\ee

\section{Strict self-adjointness}\label{classification}

Considering (\ref{3.4}) and (\ref{3.5}) one can easily conclude that the quantity (\ref{3.4}) is not a local conserved vector for (\ref{1.6}) since the components (\ref{3.4}) depend on the nonlocal variable $v$. In \cite{ib2}, see \cite{ib6,ib8} for a deeper discussion, Ibragimov introduced the concept of strictly self-adjoint differential equations. Actually, an equation (\ref{2.1}) is said to be strictly self-adjoint if and only if its adjoint equation satisfies the relation
\bb\label{3.6}
\left.F^{\ast}\right|_{v=u}=\lambda F,
\ee
for a certain function $\lambda\in{\cal A}$. If equation (\ref{2.1}) is strictly self-adjoint, then one can eliminate the non-physical variable $v$ from (\ref{2.8}) obtaining, therefore, a conserved quantity depending only on $t,\,x,\,u$ and derivatives of $u$. It means that for strictly self-adjoint differential equations, the adjoint equation is equivalent to the original one, hence the new conserved quantity is a local conserved current for the original equation and, therefore, it provides a local conservation law for it. Further details and applications of this techniques can be found in \cite{ijnmp,iamc,ijjpa,icam,icnsns,ijcnsns,gandjpa,ib3,ib4,ib5,totr}.

Let $F:=u_{t}+\eps u_{txx}+f(u)u_{x}+g(u)u_{x}u_{xx}+h(u)u_{xxx}$. Then
\bb\label{3.1}
\ba{lcl}
F^{\ast}&=&v(f'(u)u_{x}+g'(u)u_{x}u_{xx}+h'(u)u_{xxx})-D_{t}(v)-D_{x}[v(f(u)+g(u)u_{xx})]\\
\\
&&+D_{x}^2(vg(u)u_{x})-D_{x}^{2}D_{t}(\eps v)-D_{x}^{3}(vh(u)).
\ea
\ee

After a calculation of the terms involving the total derivatives in (\ref{3.1}), an explicit expression for the adjoint equation is given by
\bb\label{3.1a}
\ba{lcl}
F^{\ast}&=& -v_t -\eps v_{txx} - h(u)v_{xxx} + v(g''(u)u^3_x + 3g'(u)u_xu_{xx} - h'''(u)u^3_x - 3h''(u)u_xu_{xx}) \\ \\
& & + v_x(-f(u)-g(u)u_{xx}+2g'(u)u_x^2 + 2g(u)u_xx - 3h''(u)u_x^2 - 3h'(u)u_{xx}) \\ \\
&& +v_{xx}(g(u)u_x - 3h'(u)u_x).
\ea
\ee

Setting $v=u$ in (\ref{3.1a}) and then equaling it to $\lambda F$ to use the definition of strict self-adjointness (\ref{3.6}), we obtain
\bb\label{3.1b}
\ba{lcl}
F^{\ast}\Bigg|_{v=u}&=& -u_t -\eps u_{txx} - h(u)u_{xxx} -f(u)u_x + 
u_x^3(ug''(u) - uh'''(u) + 2g'(u) - 3h''(u)) \\ \\
&& +u_xu_{xx}(3ug'(u) -3uh''(u) - 6h'(u) + 2g(u)) \\ \\ &=& \lambda u_t + \lambda \eps u_{txx} +\lambda f(u)u_x + \lambda g(u)u_xu_{xx} + \lambda h(u)u_{xxx}. 
\ea
\ee

From (\ref{3.1b}) it is concluded, from  the coefficient of $u_{t}$, that $\lambda=-1$. From the coefficients of $u_{txx}, u_x, u_{xxx}, u_x^3, u_xu_{xx}$, respectively, one obtains

\bb\label{3.1c}
\ba{lcl}
-\eps = \lambda \eps, \quad -f(u) = \lambda f(u),\quad - h(u) = \lambda h(u), \\ \\
(ug)'' - (uh)''' = 0, \\ \\
(ug)'-(uh)'' = 0.
\ea
\ee

Integrating the coefficient of $u_xu_{xx}$ once, the condition
\bb\label{3.2}
g(u)=\f{(uh)'}{u}+\f{c}{u},
\ee
where $c$ is an arbitrary constant, is obtained. It is important to observe that the coefficients of the other derivatives in (\ref{3.1c}) do not provide any new information about the functions $f(u), g(u)$ and $h(u)$. This implies that (\ref{1.6}) is strictly self-adjoint if and only if $g$ and $h$ are related by (\ref{3.2}).

In section \ref{symmetries} we showed that equation (\ref{1.6}) forms a family of scale-invariant equations if and only if it takes the form of (\ref{1.6'}). Let us now determine under what condition the scale-invariant family of equations (\ref{1.6'}) is strictly self-adjoint. Substituting $g(u)=\sigma u^{b-1}$ and $h(u)=\delta u^b$ into (\ref{3.2}), one arrives at
$$
\sigma u^{b-1}=\delta(b+1)u^{b-1}+\f{c}{u},
$$
whose solutions are $\sigma=\delta+c$, whenever $b=0$, and $\sigma=\delta(b+1)$ and $c=0$, for $b\neq0$. Defining $\delta=-\beta$, we obtain the following family of strictly self-adjoint scale-invariant equations
\begin{itemize}
\item for $b\neq 0$, we have (\ref{1.7}) and,
\item for $b=0$, we have
\bb\label{1.7'}
u_{t}+\eps u_{txx}+\gamma u_{x} = \be \f{u_{x}u_{xx}}{u}+(\be-c) u_{xxx}.
\ee
\end{itemize}

Considering physical applications, we can assume that the constants $\be$ and $\gamma$ are positive.

As we have already pointed out, if we choose $\eps=-1$ and $\gamma=\be(b+2)$, equation (\ref{1.7}) becomes
\bb\label{1.9'}
u_{t}-u_{txx}+\be(b+2)u^{b}u_{x}=(1+b)\be u^{b-1}u_{x}u_{xx}+\be u^{b}u_{xxx}.
\ee

Note that under the change $t\mapsto\be^{-1}t$, equation (\ref{1.9'}) is then reduced to
\bb\label{1.9''}
u_{t}-u_{txx}+(b+2)u^{b}u_{x}=(b+1) u^{b-1}u_{x}u_{xx}+ u^{b}u_{xxx}.
\ee

For $b=1$, equation (\ref{1.9''}) is the well known Camassa-Holm \cite{camassa} equation, while for $b=2$ it is the Novikov equation \cite{nov}. Up to our knowledgement, it is the first time that an one-parameter family of equations connecting both Camassa-Holm and Novikov equations is reported.

\section{Local conserved currents}\label{leis}

Assuming $b\neq-2, 0$ and taking the generator (\ref{1.5}) and the components (\ref{3.4}) into account, a conserved vector for the equation (\ref{1.7}) is given by
$$
\ba{lcl}
C^{0} &=&\ds{u^{2}-\eps u_{x}^2+D_{x}\left[\f{b\eps}{3}tu_{t}u_{x}-\f{b}{b+2}\gamma tu^{b+2}+\f{2}{3}\eps uu_{x}-b \be tu^{b+1}u_{xx}-\f{2b\eps}{3}tuu_{tx}\right]},\\
\\
C^{1} &=&\ds{\f{2}{2+b}\gamma u^{b+2}+2\be u^{b+1}u_{xx}+2\eps uu_{tx}}\\
\\
&&\ds{-D_{t}\left[\f{b\eps}{3}tu_{t}u_{x}-\f{b}{b+2}\gamma tu^{b+2}+\f{2}{3}\eps uu_{x}-b \be tu^{b+1}u_{xx}-\f{2b\eps}{3}tuu_{tx}\right]}
\ea
$$

We observe that the term $D_{x}(\cdots)$ in $C^0$ along with the term $-D_{t}(\cdots)$ in $C^1$ are the components of a trivial conserved vector. Therefore, according to the discussion presented in Section \ref{conceitos}, the mentioned terms are not needed for finding a really useful conserved vector and we should simplify the established components. For this reason, after transferring the terms $D_{x}(\cdots)$ from $C^{0}$ to $C^{1}$ and eliminating the null divergence, it is obtained the components
\bb\label{6.1}
C^{0} =u^{2}-\eps u_{x}^{2},\,\,\,C^{1} =\f{2}{2+b}\gamma u^{b+2}-2\be u^{b+1}u_{xx}+2\eps uu_{tx}.
\ee

Components (\ref{6.1}), for a fixed $\eps$, provide a three-parameter family of components of conserved vectors to the family of equations (\ref{1.7}). On Table \ref{tabela1} one finds some shallow water models belonging to the family (\ref{1.7}) and their corresponding conservation laws associated with the scale symmetry $(x,t,u)\mapsto(x,\lambda^b t,\lambda u)$.

\begin{table}[h]\label{t1}\label{tabela1}
\begin{tabular}{|c|c|c|c|c|c|c|c|c|}\hline

 $\eps$    &     $b$      & $\gamma$         & $\be$   & Equation      & Conserved density & Conserved flux \\\hline

 $-1$     &     $1$      & $-1$              & $0$           & Benjamin-Bona-Mahony   & $u^{2}+ u_{x}^{2}$ & $\f{2}{3}u^3 -2uu_{tx}$\\\hline

 $-1$     &     $1$      & $3$               & $1$           & Camassa-Holm                 & $u^{2}+ u_{x}^{2}$ & $2u^3 - 2u^2u_{xx} -2uu_{tx}$  \\\hline
           
$-1$     &     $2$     & $4$                & $1$         & Novikov                              & $u^{2}+ u_{x}^{2}$ & $2u^4 -2u^3u_{xx} - 2uu_{tx} $ \\\hline

$0$     & $\neq -2$     & $\forall$                & $0$         & Riemman           & $u^2$ & $\f{2}{2+b}\gamma u^{b+2}$ \\\hline

$\forall$  &  $0$   & $\forall$  & $\forall$ & $-$   & $u^{2}-\eps u_{x}^{2}$ & $\gamma u^{2}-2(\beta-c) uu_{xx}+2\eps uu_{tx} - cu_x^2$  \\\hline

$\forall$     &     $-2$     & $\forall$                & $\forall$         & $-$          & $u^2 - \eps u_x^2$ & $-2\beta\f{u_{xx}}{u} + 2\gamma \ln u + 2 \eps uu_{xt}$  \\\hline

\end{tabular}
\caption{\small{In this table it is presented some equations of the type (\ref{1.7}) as well as some conserved currents and the corresponding symmetry from which the conserved vector was obtained. For further details about the last two cases presented in this table, see Remarks 1 and 2, respectively, below. On the penultimate line, $c$ is an arbitrary constant.}}
\end{table}

Using the same approach, some of the listed conservation laws were obtained in the last 5 years. In fact, the conservation laws for the CH equation were obtained in \cite{ib7}. In \cite{ijnmp,icam} conservation laws for Riemman equations were established using the same approach and, more recently, the conservation law for the Novikov equation was derived in \cite{bfi}. 

Although the conservation law found for the Benajamin-Bona-Mahony is well known, see \cite{bbm,ol2}, up to our knowledge it is the first time that it is obtained via the results introduced in \cite{ib7}. 

{\bf Remark 1:} Regarding the case $b=0$, it is interesting to observe that the scaling transformation $(x,t,u)\mapsto(x,t,\lambda u)$ provides a nontrivial conservation law for the equation (\ref{1.7'}).

{\bf Remark 2:} Similarly as in Remark 1, for $b=-2$ the scaling transformation $(x,t,u)\mapsto(x,\lambda^{-2}t,\lambda u)$ also gives a nontrivial conserved quantity for $$u_{t}+\eps u_{txx}+\gamma \f{u_{x}}{u^2} +\be \f{u_{x}u_{xx}}{u^3}-\be \f{u_{xxx}}{u^2}=0.$$

\section{Discussion}\label{dis}

In this paper we considered the subclasses of equation (\ref{1.6}) having two properties: strict self-adjointness and admitting the Lie point symmetry generator (\ref{1.5}). As a consequence we obtained the family (\ref{1.7}) and, after a suitable choice of the arbitrary constant, we arrived at the equivalent equations (\ref{1.9}) and (\ref{1.10}), which includes the CH, DGH and Novikov equations.

Moreover, using some recent techniques \cite{ib6,ib8} due to Ibragimov, we established conservation laws for some members of the obtained classes, as it is shown on the Table 1. It is interesting to observe that the obtained conserved quantities are those employed in the literature of completely integrable equations of the type (\ref{1.9}) to construct a first Hamiltonian for these equations. Then the results obtained in this paper suggest a connection between strict self-adjointness and integrable equations. Moreover, previous results \cite{iamc,ijjpa,ijcnsns,ib2} had shown that the KdV equation also possesses this same property.

Although there are some known examples of integrable equations that are not strictly self-adjoint, such as the Harry--Dym  (HD) and Krichever--Novikov (KN) equation, they are {\it nonlinearly self-adjoint} \cite{ib3,ib6,ib8,ijcnsns,totr}. However Ibragimov proved \cite{ib8} that, up to a multiplier, all nonlinearly self-adjoint equations are strictly self-adjoint. Moreover, it is well known that most of integrable equations are nonlinearly self-adjoint, see, for instance \cite{ijjpa,icnsns,ijcnsns}. This reinforces the suspicion of a close relationship, not clear yet, between strict self-adjointness and integrability.

On the other hand, our results also show that strict self-adjointness does not imply in the integrability, as one can easily see from the fact that the BBM equation is strictly self-adjoint, but it is not completely integrable \cite{ol2}. However, noticing that under the change $t\mapsto\be^{-1}t$, equation (\ref{1.10}) is then reduced to
\bb\label{6.2}
m_{t}-u^{b}m_{x}-(b+1)u^{b-1}u_{x}m=0,
\ee
which gives the CH equation for $b=1$ and the Novikov equation for $b=2$, but not the BBM equation. If the class (\ref{6.2}) admits more completely integrable equations, we do not know, since we are not specialists in this field. But we suspect that the answer is positive and we wait for a confirmation from well versed researchers in integrability. Moreover, we do not know, in the literature, an one-parameter family of equations unifying the CH and Novikov equations. 

Additionaly, it is not clear if, from the family (\ref{6.2}), it is possible, up to the mentioned cases, to find equations admitting either soliton or peakon solutions or when the inverse scattering method can be applied \cite{ablo}. We hope that some  enlightening about this unclear point to us could be soon presented. 

We would like to do some comparisons between (\ref{6.2}) and the $b-$equation\footnote{In fact, in the references the equation is denoted by $u_{t}-u_{txx}+(b+1)uu_{x}=bu_{x}u_{xx}+uu_{xxx}$. However, here, in order to avoid confusion, we use the form (\ref{6.3}).} (see \cite{dehoho,dugo2003})
\bb\label{6.3}
u_{t}-u_{txx}+(B+1)uu_{x}=Bu_{x}u_{xx}+uu_{xxx}.
\ee	

Clearly (\ref{6.3}) admits the scale invariance $(x,t,y)\mapsto(x,\lambda^{-1}t,\lambda u)$, since it can be obtained from the family (\ref{1.6'}) choosing $b=1$, $\gamma=B+1,\,\,\sigma=-B$ and $\delta=1$. However, comparing (\ref{6.3}) with (\ref{1.9''}) or (\ref{1.7}), we conclude that $b=1$ and $B=2$, which means that (\ref{6.3}), with these values, is the Camassa-Holm equation. 

In \cite{dehoho} it was shown that (\ref{6.3}) is integrable if $B=3$. However, such equation is not a member of our family (\ref{6.2}). In fact, comparing (\ref{6.3}) with (\ref{1.6}) it is easy to conclude that $f(u)=(B+1)u, \,\,g(u)=-B$ and $h(u)=-u$. From the condition (\ref{3.2}) we obtain
$B=2+c/u$, which implies that $B=2$ and $c=0$. Then we realise that (\ref{6.3}) is strictly self-adjoint if and only if $B=2$, which is the CH equation.

Although (\ref{6.3}) possesses among its members, the CH and the Degasperis-Procesi equation (case $B=3$), which are both integrables, it is not strictly self-adjoint, for any $B$. Moreover, these are the only integrable equations of the type (\ref{6.3}), see \cite{dehoho}. On the other hand, our new equation (\ref{6.2}) also connects at least two integrable equations, namely, Camassa-Holm and Novikov. But, differently from (\ref{6.3}), every member of (\ref{6.2}) is also strictly self-adjoint. 

Further investigations, incorporating more general classes, will be considered in forthcoming papers. Particularly we would like to consider wider classes of equations, in order to incorporate other well known equations, such as the Qiao's equation \cite{qiao}, which does not fit in the class investigated in this work.

\section*{Acknowledgements}

The authors would like to thank FAPESP for financial support (grant n. 2011/19089-6 and scholarship n. 2012/22725-4). We are grateful to Professor Valery Shchesnovich for having guided us to the book \cite{ablo}. I. L. Freire is also grateful to CNPq for financial support (grant nº 308941/2013-6).

\end{document}